# Simulation Study of Two Measures of Integrated Information

Suresh Jois[1] and Nithin Nagaraj[2]

[1]Independent Research Scholar, Bengaluru, India. suresh.jois@gmail.com
[2]Consciousness Studies Programme, National Institute of Advanced Studies, Bengaluru, India. nithin@nias.iisc.ernet.in

## Abstract

**Background:** Many authors have proposed Quantitative Theories of Consciousness (QTC) based on theoretical principles like information theory, Granger causality and complexity [1-9]. Recently, Virmani & Nagaraj [2] noted the similarity between Integrated Information and Compression-Complexity, and on this basis, proposed a novel measure of network complexity called Phi-Compression Complexity (Phi-C or $\Phi^C$). In [2] computer simulations using Boolean networks showed that $\Phi^C$ compares favorably to Giulio Tononi et al's [3-5] Integrated Information measure $\Phi$ and exhibits desirable mathematical and computational characteristics.
**Methods:** In the present work, $\Phi^C$ was measured for two types of simulated networks: (A) Networks representing simple connectivity motifs in [5: Fig 9], that according to authors of [5], are found in biological and neural systems, and that were originally studied by them; (B) random networks derived from Erdős–Rényi $G(N, p)$ graphs [10, 11]. Code for all simulations was written in Python 3.6, and the library NetworkX [12] was used to simulate the graphs described in this paper.
**Results and discussions summary: In simulations A**, for the same set of networks [5: Fig 9], $\Phi^C$ values differ from the values of IIT 1.0 $\Phi$ in a counter-intuitive manner. It appears that $\Phi^C$ captures some invariant aspects of the interplay between information integration, network topology, graph composition and entropy of nodes. While [2] sought to highlight the correlations between $\Phi^C$ and IIT $\Phi$, the results of simulations A highlight the differences between the two measures in the way they capture the essence of integrated information. **In simulations B,** the observations and discussions of simulations (A), are extended to the more general case of random networks. **In the concluding section** we outline the novel aspects of this paper, and our ongoing and future research.

## Introduction

In recent years, several Quantitative Theories of Consciousness (QTC) have been proposed. [1, 8] provide a comparative overview of QTCs. The common minimum goal of QTCs is to provide numerical measures of level of consciousness and, if allowed by the specific formulation of a QTC, contents of consciousness. QTCs formulate both the aspects—levels and contents of consciousness—as mathematical functions of properties of the underlying substrate that is presumed to possess consciousness. Alternative QTCs can be distinguished by the mathematical structures that they use for three purposes: one, to model substrate systems that can have consciousness; two, to model the level and contents of consciousness that substrates have; and three, the mathematical function, or equivalently, algorithm, that maps the substrate model to the consciousness model.

One of the main scientific goals and motivations for QTCs seems to be the need to address the Hard Problem of Consciousness [6, 7]. Alternative QTCs vary widely in their ability to capture the essence of consciousness as defined by the Hard Problem. The key issue seems to be: with what fidelity does each QTC capture the interplay, or mappings, between observable properties of substrates—such as the number of structural elements, their entropies and connectivity patterns, capability to consume and transform energy, information and other inputs—and the properties of the set of conscious experiences such as their variety, compositionality, causal power, observability and time-course? It appears that the essential difference (and by implication, veracity, consistency, tractability and testability) of different QTCs rests on the specific form that they provide to the model of substrate, the model of its consciousness and the mapping function between the two.

Many QTCs use commonly understood mathematical function*s* to map the substrate model to the model of its consciousness. These functions are based on theoretical principles like Shannon Entropy, Granger Causality, Perturbation Complexity and others, that are prevalent in several fields of science and have been used to study a wide variety of phenomena ranging from algorithms to physical systems.

The QTC 'Perturbational Complexity Index' or PCI [9] derives entirely from concrete experimental observations using a specific stimulation modality (transcranial magnetic stimulation or TMS) on a specific type of substrate: the human brain. Tononi et al's Integrated Information Theory (IIT) [3-5], also started out in its 1.0 incarnation [3] by defining a measure denoted by Phi ($\Phi$) using the well-known information theoretic concept of Mutual Information and the corresponding mathematical function. However, in the latest version IIT 3.0 [4], Tononi et al start with properties of consciousness that they deem intrinsic and exclusive to conscious experience itself—and no other phenomenon, physical or information theoretic—then specify an algorithmic procedure to compute both levels and contents of consciousness, using state transition probability distributions and their relationships within the underlying substrate. The 3.0 version of Tononi et al's QTC is still called Integrated Information Theory, but does not explicitly use any information theoretic quantity such as Entropy, Mutual Information and the like. In fact IIT 3.0 proposes a new type of information known as 'intrinsic information' which is very different from Shannon's notion of information, termed as 'extrinsic information' by the authors. In IIT 3.0 both substrate and consciousness are abstract structures and can model any physical entity, be it atoms or neuronal networks, and their corresponding conscious experiences.

QTCs like PCI [9] are empirically derived, hence tractable for experimental, theoretical and simulation explorations. IIT 3.0 has a simple, if abstract, definition of substrate, but a relatively complex definition of consciousness and the mapping function. The authors themselves in [4] discuss challenges of computational and experimental verification arising from their formulations of the models of substrate, consciousness and the mapping function.

Recently, Virmani & Nagaraj [2] noted the correlation between Integrated Information and Compression Complexity, and on this basis, proposed a novel measure of integrated information called Phi-Compression Complexity (Phi-C, or $\Phi^C$).Their QTC includes a model for the level, but not contents of consciousness. The procedure for computing $\Phi^C$ is an algorithmic analog of PCI's [9] experimental procedure. Based on computer simulations of boolean networks, the authors of [2] claim that $\Phi^C$ is computationally more tractable than IIT-$\Phi$, that it compares favorably to IIT-$\Phi$, and that it exhibits desirable mathematical and computational characteristics.

Given the variety of QTCs extant in literature, there is a strong need for objective and quantitative comparisons between them for the same types of substrates, be they neuronal networks, connectivity motifs, graphs or other types of structures. The present paper is an effort in this direction. The present paper extends the work of [2] using two simulation studies:

**In Simulations A,** for the same set of networks as in [5: Fig 9], we compared $\Phi^C$ with IIT 1.0 $\Phi$. We find that $\Phi^C$ values differ from the values of IIT 1.0 $\Phi$ in a counter-intuitive manner. It appears that $\Phi^C$ captures essential and invariant aspects of the interplay between information integration, network topology, node entropy and graph composition. While [2] sought to highlight the correlations between $\Phi^C$ and IIT 1.0 $\Phi$, the emphasis of this paper is on the subtle differences between the two measures in the way they capture the essence of integrated information and dynamical complexity of networks.

**In Simulations B** the observations and discussions of simulations (A), are extended to the more general case of random networks derived from Erdős–Rényi *G(N, p)*[10,11] graphs.

The rest of the paper is organized into four sections. The next *Methods* section summarizes the methods and assumptions common to all simulations. This is followed by two sections, one each per simulation study (A) and (B) mentioned in the previous paragraph. Within each section, a Results subsection contains a summary of the outputs from the respective simulation; a Discussion subsection contains the main observations, inferences and hypotheses drawn from the simulation results. The final section provides concluding remarks on the entirety of this paper, and an outline of ongoing and future work to extend the results of this paper.

# Methods

All simulations were performed using the Effort-to-Compress (ETC) formulation of $\Phi^C$ described in [2]. The simulation procedure for $\Phi^C$ itself is described in detail in [2]. As such, it is not repeated here. Code for the simulations in this paper was written in Python version 3.6 on a machine running Microsoft Windows 7. Network topologies targeted for simulation were realized using the Python library NetworkX [12]. Maximum Entropy Perturbation (MEP) and Zero Entropy Perturbation (ZEP) inputs for the simulated networks were generated in the Python code, and outputs from the simulated networks were fed into $\Phi^C$ computation function written in Matlab.

In [2], simulations were performed on Boolean networks, and sensitivity of $\Phi^C$ to different distributions of initial state of the network was an important object of study. In this paper, however, study of initial state sensitivity was not an objective. As such, all simulations were performed with initial state of all network nodes set to zero. In the text of this paper, for the sake of brevity, values are not provided for number of bins used to compute ETC based $\Phi^C$, perturbation time series used to compute $\Phi^C$, and other common parameters described in [2], since the results and discussions presented here are not dependent on them.

## Simulations A: $\Phi^C$ in networks with simple connectivity motifs

In Figure 9 of [5], IIT 1.0 $\Phi$ is computed for five types of digraphs that the authors of [5] deem "… constitute frequent motifs in biological systems, especially in neural connection patterns." The five digraphs from Figure 9 of [5] are repeated in Figure 1 below.

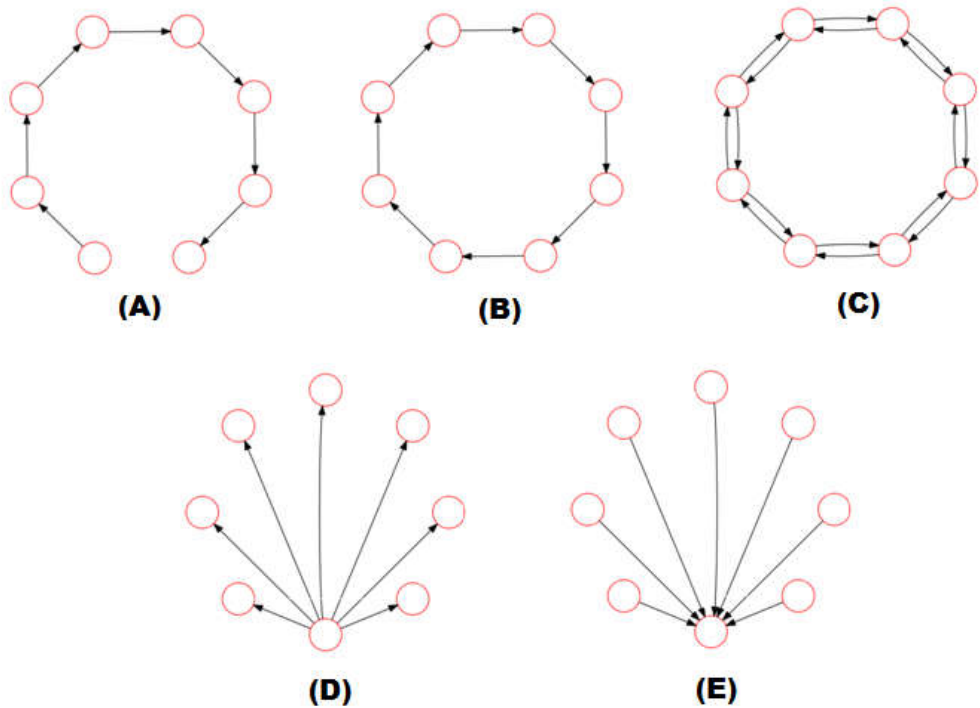

**Figure 1.** Digraphs from [5: Fig. 9]. **(A)** Simple Path, **(B)** Clockwise Cycle, **(C)** Bidirectional Cycle, **(D)** Fan-out Star, **(E)** Fan-in Star.

# Results of Simulations A

In this paper, $\Phi^C$ and IIT 1.0 $\Phi$ were computed for the same 8-node digraphs as in [5: Fig. 9]. The outputs from these simulations are as summarized in Table 1 and Figure 2 below.

**Table 1.** Comparison between IIT 1.0 $\Phi$ and $\Phi^C$ for 8-node digraphs shown in Fig. 1.

| 8-node digraph type | IIT 1.0 -$\Phi$ | $\Phi^C$ |
|---|---|---|
| A. Simple path | 10.1286 | 4.3266 |
| B. Clockwise cycle | 20.2533 | 4.3266 |
| C. Bidirectional cycle | 40.5065 | 4.3266 |
| D. Fan-out star | 10.8198 | 4.3266 |
| E. Fan-in star | 10.8198 | 0.6181 |

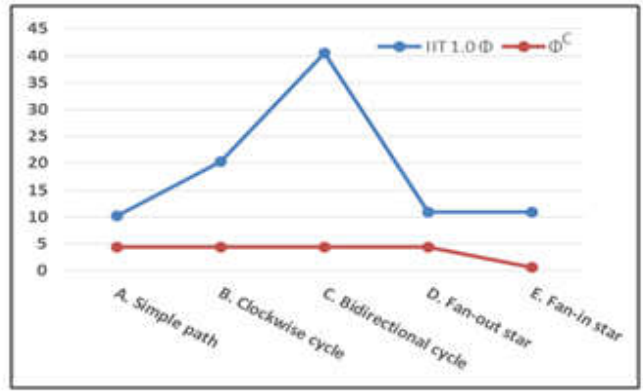

**Figure 2.** Integrated Information measure IIT 1.0 $\Phi$ vs. Compression-Complexity measure $\Phi^C$ for 8-node digraphs.

# Discussion on simulations A

As can be seen from Table 1, for the same set of digraphs, the distribution of $\Phi^C$ values is very different from that of IIT 1.0 $\Phi$. From the above results, the following main observations and hypotheses can be derived:

***IIT 1.0 $\Phi$ is different for digraphs A-D of Table 1, but $\Phi^C$ is the same for them.***

All four digraphs are connected, and entropies of all nodes in the digraphs are the same. Hence the amount of information integration achieved should be the same for all of them, regardless of differences in topology of the graphs. The distribution of $\Phi^C$ across digraphs A-D seems to conform to this principle. Indeed, in [2], all simulated Boolean digraphs were also connected, but node entropies varied between digraphs. This yielded different values of $\Phi^C$ between, for example, an AND-AND-AND graph, and an AND-OR-XOR graph.

***IIT 1.0 Φ ranks the first three digraphs of Table 1 in the order C > B > A, but $\Phi^C$ is the same for all three.***

IIT 1.0 Φ value for digraphs A, B, C seems to be proportional to the number of paths or edges in the digraphs. But $\Phi^C$ seems to depend only on whether the digraphs are connected, not on the number of edges or paths in them.

***IIT 1.0 Φ is the same for digraphs D, E, but $\Phi^C$ of digraph E is a fraction of that of D.***

In digraph D, the hub node is sending the same information to all other nodes, so all nodes effectively possess the same information about each other. In digraph E, the hub node is receiving information from other nodes but not vice versa, so the non-hub nodes do not possess information about the hub node or about each other. This entails that even though both graphs are connected, functionally the potential for information integration is higher in digraph D than in E. $\Phi^C$ reflects this whereas IIT 1.0 Φ fails to distinguish the two cases.

***IIT 1.0 Φ value for digraph C is double that of B, while $\Phi^C$ is the same for both. IIT 1.0 Φ and $\Phi^C$ behave differently under graph composition.***

Digraph C is a union of cycle digraph B and its reverse graph. If graph union does not create entirely new patterns of effective connectivity between nodes compared to that of its input graphs, then the amount of integrated information should remain unchanged under graph union. $\Phi^C$ seems to conform to this.

**In summary, $\Phi^C$ captures the potential for information integration in networks, taking into account both their physical connectivity and functional connectivity.**

# Simulations B: $\Phi^C$ for random digraphs and under graph unions

In Simulations A, it was noted that $\Phi^C$ has the same value for a cycle digraph, its reverse graph and the union of the two. To corroborate this behavior in a more general setting, $\Phi^C$ was computed for the more general case of random networks derived from Erdős–Rényi *G(N, p)* [10,11] graphs that had the same number of nodes=8, as the digraphs in Simulations A, but with varying edge creation probability. The graphs in Simulations B are depicted in Figure 3.

The objective of simulations B was to observe the change in value of $\Phi^C$ as the edge-creation probability of graphs was varied from a very low value corresponding to a sparse graph, to a value of 1.0 corresponding to a complete graph. $\Phi^C$ was computed for each digraph, its reverse digraph and union of the two digraph. The results are summarized in Table 2.

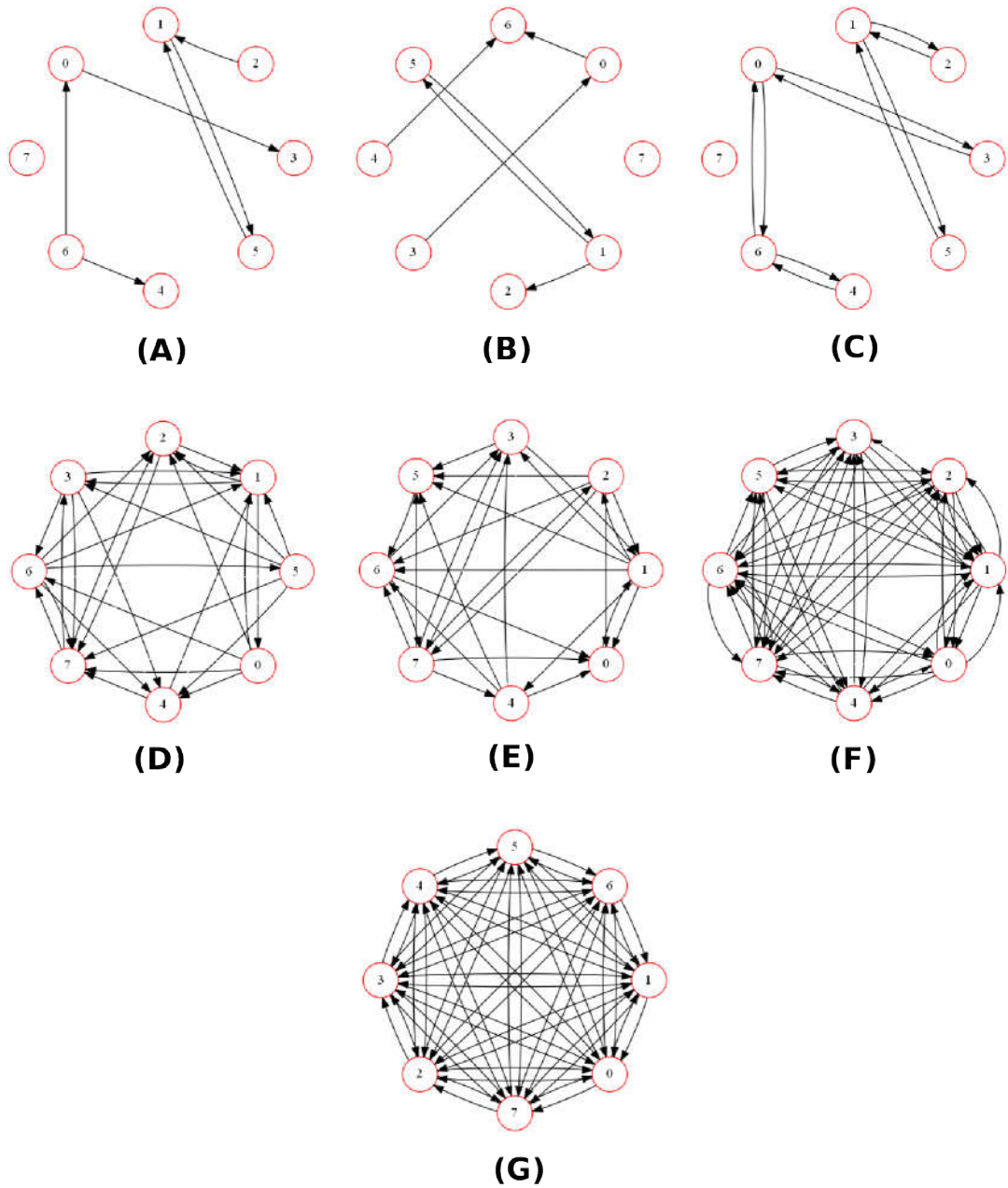

**Figure 3.** Erdős–Rényi *G(N, p)* random graphs used in Simulations B. **(A)** Edge creation probability *p* = 0.1, **(B)** Reverse graph of (A), **(C)** Graph (A) ∪ (B), **(D)** *p* = 0.53, **(E)** Reverse graph of (D), **(F)** Graph (D) ∪ (E), **(G)** For edge creation probability *p* = 1.0, the random graph, its reverse and union of the two are all identical, complete graphs.

## Results of simulations B

Table 2. $\Phi^C$ values from simulations B

| Edge creation probability → | p=0.1 | p=0.53 | p=1.0 |
|---|---|---|---|
| $\Phi^C$ for 8 node random digraph with p | 1.8693 | 4.3618 | 4.3618 |
| $\Phi^C$ for reverse digraph of above digraph | 1.2462 | 4.3618 | 4.3618 |
| $\Phi^C$ for union of above two digraphs | 1.8693 | 4.3618 | 4.3618 |

## Discussion on simulations B

From the above results, the following main observations and inferences can be made:

***$\Phi^C$ is invariant under digraph union***.

This was observed in Simulations A for the specific case of Cycle digraph B, and C which is the union of B and its reverse digraph. The same can be observed in the above results for random graphs of varying edge-creation probability.

***$\Phi^C$ for edge creation probability above a certain threshold is the same as for complete graphs***.

Random digraphs with edge creation probability $p$ equal to or above a certain threshold (in this instance 0.53) behave like connected graphs in regards to integrated information, and of course $p = 1.0$ yields complete graphs. Expectedly, for these digraphs, their reverse digraphs and the unions of the two, $\Phi^C$ remains invariant.

***$\Phi^C$ is unequal for a random digraph and its reverse digraph for low edge creation probability.***

For edge creation probability $p = 0.1$, there is a reduction in $\Phi^C$ going from the initial random digraph and its reverse. This effect needs to be studied further across multiple trials of low $p$.

## Concluding remarks

In Simulations A and B, we studied the distribution of $\Phi^C$ in digraphs representing simple connectivity motifs and random graphs. In all cases, the results yielded interesting new behavior of $\Phi^C$ in qualitative and quantitative terms. Based on these results, in the preceding sections we posited several observations and hypotheses regarding the analytical and phenomenal nature of $\Phi^C$, both intrinsic to itself and in comparison to IIT 1.0 $\Phi$.

In the introduction section, we observe that QTCs are based on diverse conceptual, physical and computational principles. Also QTCs are at diverse states of maturity in regards to measure and contents of consciousness. Empirical verification of QTCs in in-vivo, in-vitro and in-silico neuronal substrates of realistic biological plausibility faces challenges of computational complexity as well as of functional and connectomic data. These issues indicate that mutual comparative study of QTCs on an objective quantitative basis is also a challenging task, but all the same, a necessary one. In this paper we have made a beginning in this direction. In this section, we outline the potentially novel aspects of this paper, and also ongoing and future work in continuation of the work described in this paper.

## Potentially novel aspects of this work

There have been attempts to characterize and compare QTCs in works such as [1, 8]. However the predominant approaches deployed have been descriptive and conceptual classification oriented. Quantitative and objective comparison of QTCs remains a sparse and challenging endeavor due to the issues briefly mentioned in the previous paragraph.

This paper represents an attempt to compare two QTCs by these quantitative methods:
- Using the same simulated network motifs.
- Taking a graph theoretic approach to characterize and compare QTCs, in this instance
  - With the functional principle of network graph composition operation.
  - Using random graphs with varying topology.

To the best of our knowledge, this is a novel attempt wherein the above techniques have been deployed in conjunction for objective comparison of QTCs of different conceptual flavors. Further, in respect of $\Phi^C$ itself, we present new results on its distribution and behavior in the types of network topologies described in this work.

## Ongoing and future work

From its present stage of specification, $\Phi^C$ would need to evolve into a more comprehensive QTC, especially to describe contents of consciousness. A QTC based on $\Phi^C$ would need to comprehensively specify what aspects of consciousness it actually captures, both phenomenologically and analytically. There is also a need to relate the analytical and computational properties of $\Phi^C$, to a mathematically based phenomenological scheme, as has been done by Tononi et al with their concept of qualia space [4].

The present authors are engaged in the following activities to accomplish the above goals:

- Mathematically formalizing the observations and hypothesis stated in this work.
- Extend the graph theoretic approach used to characterize QTCs in this work, to other types of graph operations.
- Relate observations and hypothesis in this work to phenomenology and formal properties of data compression.
- Relate observations and hypothesis in this work to phenomenology and formal properties of information theoretic quantities like Shannon Entropy.
- Relate observations and hypothesis in this work to phenomenology and formal properties of physical quantities and phenomena like Energy and Causality.

- Objective and quantitative comparison of $\Phi^C$ with other measures of brain complexity and consciousness besides IIT.
- Extend this work to functional networks derived from clinical data of normal brain states and neurological disease conditions such as epilepsy.
- Extend this work to other types of network connectivity motifs such as small world networks, scale-free networks and so on.
- Evolve a comprehensive phenomenology for $\Phi^C$ along with $\Phi^C$ derived measures for contents of consciousness and their dynamics.

**Note:** The data and discussions presented in this paper are based on single trial simulations. They are meant only for dissemination of key conceptual ideas related to $\Phi^C$ on different network topologies.